\documentclass[suppldata]{interact}
\usepackage{epstopdf}
\usepackage[caption=false]{subfig}
\usepackage[numbers,sort&compress]{natbib}
\bibpunct[, ]{[}{]}{,}{n}{,}{,}
\renewcommand\bibfont{\fontsize{10}{12}\selectfont}
\makeatletter
\def\NAT@def@citea{\def\@citea{\NAT@separator}}
\makeatother

\usepackage[usenames,dvipsnames]{color} 
\usepackage{float}
\usepackage{xltabular}
\usepackage[textsize=tiny]{todonotes}
\usepackage{hyperref}

\newcommand{\invitro}{\emph{in vitro}}%
\newcommand{\INVITRO}{\emph{In vitro}}%
\newcommand{\invivo}{\emph{in vivo}}%
\newcommand{\INVIVO}{\emph{In vivo}}%
\newcommand{\exvivo}{\emph{ex vivo}}%
\newcommand{\insilico}{\emph{in silico}}%
\begin{document}
\articletype{REVIEW ARTICLE}
\title{Recipes for calibration and validation of agent-based models in cancer biomedicine}
\author{
\name{
Nicol{\`o} Cogno\textsuperscript{a}, 
Cristian Axenie\textsuperscript{b},
Roman Bauer\textsuperscript{c}
\thanks{Email: \email{r.bauer@surrey.ac.uk}} and
Vasileios Vavourakis\textsuperscript{d,e}
\thanks{Email: \email{vavourakis.vasileios@ucy.ac.cy}}
}
\affil{
\textsuperscript{a} Institute for Condensed Matter Physics, Technische Universität Darmstadt, 64289 Darmstadt, Germany; \\
\textsuperscript{b} Computer Science Department and Center for Artificial Intelligence, Technische Hochschule N{\"u}rnberg Georg Simon Ohm, N{\"u}rnberg, Germany; \\
\textsuperscript{c} Department of Computer Science, University of Surrey, Guildford, UK;\\
\textsuperscript{d} Department of Medical Physics and Biomedical Engineering, University College London, London, UK; \\
\textsuperscript{e} Department of Mechanical and Manufacturing Engineering, University of Cyprus, Nicosia, Cyprus }
}
\maketitle
\begin{abstract}
Computational models and simulations are not just appealing because of their intrinsic characteristics across spatiotemporal scales, scalability, and predictive power, but also because the set of problems in cancer biomedicine that can be addressed computationally exceeds the set of those amenable to analytical solutions. Agent-based models and simulations are especially interesting candidates among computational modelling strategies in cancer research due to their capabilities to replicate realistic local and global interaction dynamics at a convenient and relevant scale. Yet, the absence of methods to validate the consistency of the results across scales can hinder adoption by turning fine-tuned models into black boxes. This review compiles relevant literature to explore strategies to leverage high-fidelity simulations of multi-scale, or multi-level, cancer models with a focus on validation approached as simulation calibration. We argue that simulation calibration goes beyond parameter optimization by embedding informative priors to generate plausible parameter configurations across multiple dimensions.

\end{abstract}
\newpage

\section{Agent-based modelling in biomedicine}

\begin{figure}[H]
     \centering
     \includegraphics[width=0.61\textwidth]{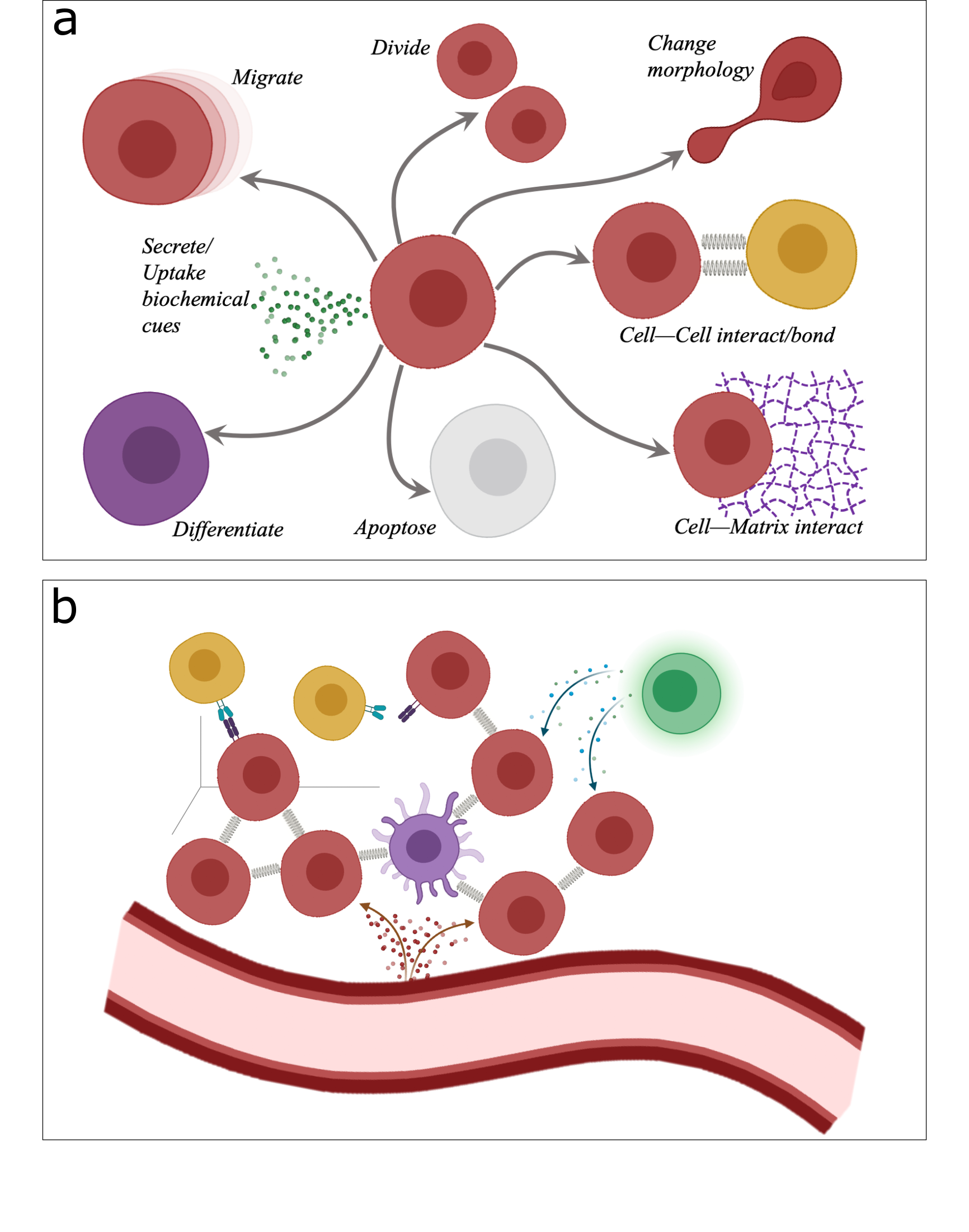}
     \caption{Autonomous cell agents interacting with each other in an ABM simulation.
     (a) A common (not exhaustive) set of cell behaviour mechanisms includes: `Divide,' `Migrate,' `Change morphology' (grow, shrink, polarize), `Die' (programmable, stimulated), `Differentiate,' `React to a chemical cue,' `Secrete a substance,' and `Interact' with other cells or the extracellular matrix. (b) The ABM software enables the simulation of the model dynamics. }
     \label{fig:1}
\end{figure}

Agent-based modelling (ABM) is a relatively recent mechanistic numerical procedure with biological applications \cite{Gary_et_al:2009, Wilensky:2015} that represents processes and phenomena in terms of computational agents. 
Agents can be denoted as particles to represent cells in biology, or segments to represent neurites or vessels, which can reside in space either in a structured lattice or following an off-lattice approach.
In ABM, agents (cells) are programmed with respect to their behaviour and interaction with other agents, modelled as Markov processes using mathematical rules for their description. 
When the agents' decisions arise from probabilistic reasoning, stochastic systems can be simulated and complex, higher-scale behaviours emerge as the simulation clock ticks. 
The agents' rules can, under suitable conditions, enable multi-cellular systems to self-organise into highly non-random structures.
Thus, ABM simulations in cell biology are characterised by the dynamics of autonomous and heterogeneous entities whose phenotypic behaviour is explicitly modelled (see Fig.~\ref{fig:1}a), yet cell behaviour can adapt in time based on cell--cell and cell--microenvironment interactions (see Fig.~\ref{fig:1}b).

\subsection{Mechanistic models in ABM (2010--present)}
\label{sec:Mechanistic_ABM}

Intrinsically decentralised, inherently interactive, and multi-entity, biological systems are ideal to simulate using ABM.
Among other diseases, cancer biology has witnessed an increasing number of modelling efforts via ABM over the years;
ABM, alone or coupled with other \insilico{} modelling techniques, has been employed to explore the dynamics of the tumour microenvironment (TME), to probe and identify new therapeutic agents, to promote clinical translation by aiding the development of new diagnostic tools, and to bridge the gap between animal and human data \cite{Clancy_et_al:2016, Rockne_et_al:2019, Hadjicharalambous_et_al:2021}. 
The short survey of Metzcar et al.~\cite{Metzcar_et_al:2019} presents the state-of-the-art ABM simulations related to cancer hypoxia and necrosis, tumour-induced angiogenesis, invasion, stem cell dynamics and immuno-surveillance in neoplasia during the 2000s and the first half of the previous decade.
In this first section, a survey of the major agent-based approaches of the last decade to model cancer development and potential therapeutic strategies is presented.

\begin{figure}[H]
    \centering
    \includegraphics[width=0.75\textwidth, angle=-90]{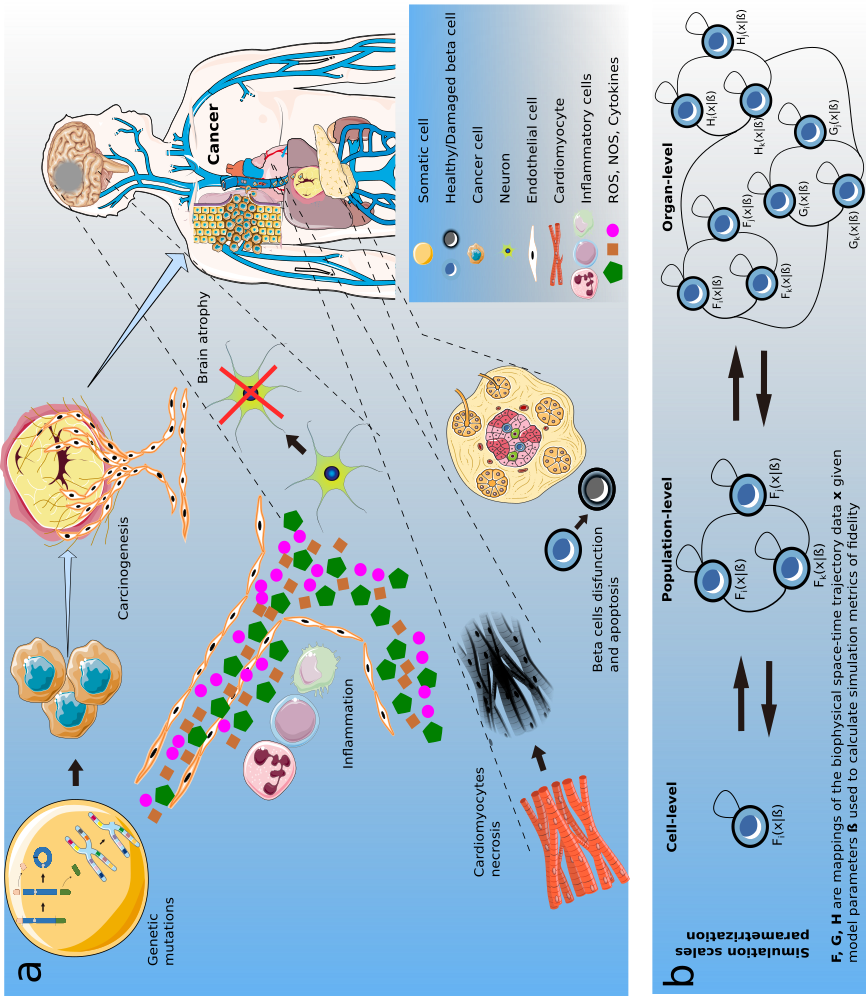}
    \caption{a) Diseases pathogenesis. Environmental factors, ionising radiation, and an unhealthy lifestyle (e.g.\ smoke, obesity) are often regarded as triggers for the pathogenesis of, among others, cardiovascular and neurodegenerative diseases, cancer and diabetes. Together with chronic inflammation, where inflammatory cells (macrophages, T lymphocytes and neutrophils) secrete large amounts of reactive/nitrogen oxygen species and cytokines, these factors can impair the normal functioning of cells. As a consequence, apoptosis and necrosis of cardiomyocytes, beta cells in the islets of Langerhans and neurons can lead to cardiomyopathy, diabetes, and Alzheimer’s disease, respectively. The aforementioned risk factors can also cause genetic mutations and chromosome instability in somatic cells which can, in turn, switch to a hyperplastic, malignant phenotype and become cancerous. As these cells proliferate and consume nutrients, carcinogenesis is initiated, which stimulates vasculogenesis and tissue remodelling. Finally, if circulating tumour cells extravasate the blood vessels' epithelium, other organs can be invaded. b) Calibration process formalism of multi-scale biophysical processes. From cell-level homeostatic dynamics, to competition and cooperation at population level, and up to organ-level phenotypic interactions.}
    \label{fig:2}
\end{figure}

\subsubsection{Pathogenesis}

Despite being just nodes in a much broader network, chronic inflammation, unhealthy lifestyle, and environmental factors have long been regarded as triggers in the pathogenesis of cancer (see Fig.~\ref{fig:2}a). 
An et al.~\cite{An_et_al:2015} introduced an inflammation and cancer development ABM platform, built on a hierarchy of relationships between cancer hallmarks, where higher-order processes are driven by lower-order ones. 
The basal DNA damage rate of healthy cells was exacerbated by reactive oxygen species secreted by inflammatory cells (e.g., neutrophils, monocytes, and macrophages) and, if the DNA-repair rate was exceeded, an abstract genome was impacted and cells’ functions were impaired. 
Damaged cells can also recruit more inflammatory cells, resulting in the establishment of a positive feedback loop that highlights the potential role of anti-inflammatory therapies in cancer care. 
Thus, in \cite{An_et_al:2015}, cancer was presented as an evolutionary process, and an interesting analogy between oncogenesis and evolution was highlighted, both requiring a basal level of genetic instability. 
Contextually, \cite{Araujo_et_al:2013} and \cite{Lynch_et_al:2022} emphasized on the role of genomic instability in carcinogenesis. 
In the first model, the impact of each gene was proportional to the number of copies in the genome, where abstracted genes were used for division, death, and segregation mechanisms. 
When in homeostasis, the effects of proto-oncogenes (linked to the cell growth) and tumour suppressor genes were balanced, but cell duplication might randomly lead to aneuploidy in daughter cells causing hyperplastic growth. 
Different treatments were simulated, which eventually led to the conclusion that the best prognosis arises from a combination of chemotherapy and surgery. 
Besides, the simulations generated novel genotypes that resemble patterns found in cancer patients. 
Similarly, the model presented in the work by Lynch et al.~\cite{Lynch_et_al:2022} linked chromosomal instability (CIN) with the cells’ karyotype. 
A framework to measure the CIN level was built and three models for the selection of a cell after the division were compared. 
Following the division, the two cells underwent a selection procedure and, if not fit, they are removed from the agent-based simulation. 
The total fitness is then computed as a sum of the single chromosomes’ contributions and phylogenetic trees were used in combination with approximate Bayesian computation to estimate the probability of missegregation from an observed population of cells. 
The results showed that sampling karyotypes in a cell population don't allow direct determination of the CIN, as other factors play key roles, while selection and missegregation shape the karyotype diversity in a population of aneuploid cells. 

\subsubsection{Neoplastic cell dynamics}

Cells in neoplasia can undergo phenotypic switching multiple times during their lifetime which is typically driven by endogenous or/exogenous factors, such as the lack of nutrients or mechanical stimuli.
The ability to change their behaviour and adapt to the surroundings can be effectively modelled by agent-based models, as demonstrated for example in the paper of Chen et al.~\cite{Chen_et_al:2015}. 
They simulated the avascular growth of an \invitro{} tumour spheroid via a two-dimensional (2D) agent-based model, where nutrients are supplied solely by the existing environment. 
Interestingly, some modelling concepts were borrowed from the evolutionary game theory in that phenotype switching (mostly proliferation--migration and vice versa) is regulated via a payoff matrix.
The phenotype-to-phenotype competition was modelled as a rewarded game where the environment (i.e., the available resources) influences  cells’ phenotype as well as the tumour rate of growth and the surface roughness, which can be quantified. 
In Kareva \cite{Kareva_et_al:2016} and Phillips \cite{Phillips_et_al:2020} and their colleagues,  vascular tumour growth was simulated using ABM driven by tumour hypoxia and the subsequent secretion of pro-angiogenic cytokines (e.g., vascular endothelial growth factor). 
The models showed that the rate of diffusion and consumption of the growth factors is key in the angiogenesis and the competition game is reiterated, although in this case, it concerns high and low-affinity angiogenesis regulators. 
Reducing the tumour--endothelium communication would therefore allow control over the vasculogenesis mechanisms as shown, for instance in Wang et al.~\cite{Wang_et_al:2013}, where the Loewe combination index \cite{Zhao_et_al:2004} revealed that merging cell-killing drugs and vascular endothelial growth factor inhibitors provides the best treatment for melanoma. 
These mechanisms were further investigated in the work of Lima and his colleagues \cite{Lima_et_al:2021}, where a novel coarse-grained modelling approach was introduced. 
Multiple cells with the same phenotype were modelled using a single agent and the total computational time of the agent-based model dropped by 93-97\% with little difference with respect to the cell agents model. 
A multi-scenario (i.e. with different initial conditions) calibration of the model was performed and they performed moment-based Bayesian inference to calibrate their agent-based model. 
Finally, a time-dependent global sensitivity analysis allows the authors to identify the increase in the death rate due to lack of glucose as the most influential parameter. 
Overall, all spheroid model simulations showed that, eventually, the core of the tumour was dominated by necrotic cells, while the ones alive were located at the rim of the tumour.

Predicting cell heterogeneity and understanding the relative impact of intrinsic versus environmental factors in its emergence is of paramount importance; 
this is often seen as the most influential factor in multi-drug resistance, treatment failure and relapse. 
Although not feasible through \invivo{} models, \insilico{} models provide frameworks to quantitatively measure these relative weights. 
In the paper \cite{Gallaher_et_al:2020}, the authors developed a model of intra-tumour heterogeneity in glioblastoma (GBM) and used the \insilico{} model to show that some level of intrinsic heterogeneity is required to capture the migration behaviour observed in single-cell data, while the environmental heterogeneity alone was insufficient. 
In their model, GBM growth was driven by the platelet-derived growth factor that acted both in a paracrine and autocrine manner.
Combining treatments that inhibit cancer cell proliferation and suppress cell invasiveness, i.e.\ migration, led to an increased efficacy with respect to monotherapies. Moreover, the model not only proved the predictive capacity of single-cell data \insilico{}, but it also emphasized its importance by showing that cell populations with heterogeneous phenotypes displayed similar growth dynamics and final density distributions. 
Interestingly, Greene and his colleagues \cite{Greene_et_al:2015} further explored the role of cell heterogeneity on tumour growth using a continuous-time Markov chain model to describe the transitions in cell state and constrained trust region algorithms with nonlinear least squares for parameter estimation. 
The phenotypic trade-off was similarly investigated in paper \cite{Gerlee_et_al:2012} that focused on GBM, and later in \cite{Gallaher_et_al:2019}. Here, the migration (favoured by selection at the early stages of tumour development) and the proliferation (favoured more in the later phases) capabilities improved simultaneously up to a certain coordinate of the trait space, from where the increase in one of the rates could only come at the expense of the other. 
Different shapes of the trait space were analysed and an inverse proportionality between the rate of cell turnover and phenotypic variability among cancer cells was found. 
Modelling of GBM has been also the focus of \cite{Grimes_et_al:2020}, where special emphasis was put on the interplay between hypoxia and cancer progression. 
The results of their study illustrated differences in terms of the spatial distribution of oxygen/nutrients within the TME, which in turn can affect not only the rate of growth of the carcinoma but also the migrational capacity of the glioma cells. 
Notably, the authors draw attention towards hypoxia as a catalyst for dangerous mutations at a higher rate. 
In an effort to model the phenotypic transitions of cancer cells, Axenie and Kurz \cite{Axenie_Kurz:2020} proposed a model that learnt the mechanistic rules governing cancer's cells phenotypic staging from quiescent to proliferation and from proliferation to apoptosis. 
Using a typical cancer phenotypic state space, quiescent cancer cells (Q) can become proliferative (P) or apoptotic (A). 
Non-necrotic cells become hypoxic when oxygen drops below a threshold value, while hypoxic cells can recover to their previous state or become necrotic. 
The transitions among these states are stochastic events generated by Poisson processes. 
Although trained on a limited time series of raw immunohistochemistry and morphometric data from 17 patients, the lightweight machine learning system was also able to accurately predict tumour volume evolution and the dependency between histopathological and morphological data, such as nutrient diffusion penetration length within the breast tissue, the ratio of cell apoptosis to proliferation rates and radius of the tumour.
More recently, Gazeli and his colleagues \cite{Gazeli_et_al:2022} employed ABM to simulate \invitro{} experiments on melanoma (B16F10) cancer cells monolayer growth when treated with Doxorubicin alone, or in combination with treatment based on cold atmospheric pressure plasma jet.
Their model was designed in order to probe the mechanisms of action of each therapeutic approach (cytotoxic drug or/with plasma); this was characterised through model-derived probabilities of the melanoma cell apoptosis and division.
They presented an interesting approach that combines \insilico{} with \invitro{}, and they demonstrated how simulations can help to speed up laboratory work and, thus, reduce the costs for cancer drug/treatment testing.

An interesting finding is reported in Poleszczuk et al.~\cite{Poleszczuk_et_al:2014}, where the authors observed a reduction in the tumour volume by exploiting the induction of cancer cells' senescence.
They presented an ABM procedure to probe the competition of cell sub-populations where cancer-stem cells were allowed to ``fight'' against cancer progenitor cells to ensure vital space and nutrient supply. 
Their simulations showed that the tumour growth is regulated in two distinctive phases, in that an initial increase of the cancer cell population was followed by a reduction in the proliferation rate and eventually tumour control. 
The latter was in fact caused by the reactivation of the senescence program in the progenitor cells that constrained the stem cells in the tumour core, thereby limiting their proliferation. 
Similar \insilico{} results concerning neoplasia spatial inhibition were presented by Norton et al.~\cite{Norton_et_al:2014}; their focus was to investigate the role of cancer cell seeding in metastatic tumour progression. 
In particular, the authors modelled the metastatic cells dissemination in two potential scenarios: the `site' seeding, where cancer-stem cells were injected from a single direction due to a breach in the vasculature, and the `volume' seeding, where seeding was allowed to happen at random locations in the metastatic tumour region. 
While migration promoted tumour growth in every scenario, even when the cell division rate was considered high, volume seeding enhanced tumour growth. 
However, the impact of the seeding procedure was reported higher when cells’ quiescence inhibited spatial growth. 
Moreover, by extending the simulator into a three-dimensional (3D) ABM simulator, the authors were able to recapitulate the visual differences in the tumour morphology with respect to the proliferation and migration parameters of their stochastic model. 
Recently, de Montigny and his colleagues~\cite{deMontigny_et_al:2021} proposed a hybrid approach that uniquely combines agent-based and finite element modelling to simulate GBM progression and bridge the gap between continuum-based and discrete system dynamics. 
While the transport of nutrients and intra-cellular signalling was simulated using the finite element method (FEM), the cells were modelled as agents, with volume averaging used to interface the two spatial scales (tissue and cell scale).
Their methodology helped reduce by several orders of magnitude the number of simulated agents and, consequently, bring down the total simulation time. 
Notwithstanding this, the hybrid model can replicate growth patterns of both low- and high-grade tumours and assesses the role of platelet-derived growth factor on the tumour shape and size at later time points. 
A somewhat similar methodology was presented by Rahman et al.~\cite{Rahman_et_al:2017}, where GBM growth was modelled at multiple spatial and temporal scales, ranging from sub-cellular signalling pathways to the progression of the tumour tissue. 
By combining PDE solvers, an ABM simulator and ODE solvers for the tissue, cellular and inter-cellular scales, respectively, the authors provided a coupled cancer model where inter-compartmental communication ensures synchronization. 
The model was tested to replicate experimental findings concerning tumour growth and cell proliferation both in physiological conditions and following the administration of biochemical compounds (rapamycin).

\subsubsection{Spatial characterisation}

Tumour diagnosis together with the best treatment strategies can benefit from the characterisation of the shape and spatial features of a solid tumour.
Structural imaging, such as magnetic resonance imaging (MRI) and histology images, and \insilico{} modelling, such as ABM simulations, can provide spatial information and make predictions of the geometric features of carcinoma as they develop over time. 
Focusing on ABM, local modifications in the TME can be linked to outcomes at the tumour scale, overcoming the limitations imposed by reaction-diffusion equations that take average parameters from MRI or Computed-Tomography scans. 
As argued by Klank and her colleagues \cite{Klank_et_al:2018}, modelling GBM growth as a Brownian motion via mechanistic rules of an agent-based formulation is ample to replicate the development of a highly packed tumour core with an enhancing (proliferative) boundary at the tumour--host interface, where the overall expansion speed of the lesion depends on single-cell migration rates. 
In Norton \cite{Norton_et_al:2017} and Karolak \cite{Karolak_et_al:2019}, shape metrics such as the mean chord length, the moment of inertia, the radius of gyration and the accessible surface area are used to characterise tumour morphology and packing density.
The results show that this data can be extracted from diagnostic images and allow for tumour invasiveness and cancerous nature predictions.
And while the more generic and commonly used tumour diameter doesn't provide information about the tumour architecture, the aforementioned metrics could be adopted to supplement this limitation. 
In fact, at similar tumour sizes substantial morphological differences may be concealed. In this regard, computational models could be used to map anatomical compositions (in terms of shape metrics) to the corresponding effective drug penetration rates. 
These, in turn, could be employed to adapt the therapeutic protocols (e.g., drug doses and schedule) prior to treatment. 
Moreover, Norton et al.~\cite{Norton_et_al:2017} reported that tumour morphology and invasiveness are directly linked to the cancer cell phenotypic ratios, the level of hypoxia and the number of chemokine receptors. 
In fact, reduced tumour growth was observed following the virtual administration of a drug that impaired cancer cell proliferation. 
Similarly, Bull et al.~\cite{Bull_et_al:2020} modelled via ABM the advective flow of microspheres from the tumour rim to the tumour core that results from the outer pressure in tumour spheroids. 
Their simulations illustrated distinct spherical shells in which the cells’ movement was either dominated by Brownian motion or advection. 
More specifically, the diffusion of the micro-beads located in the tumour rim was Brownian-dominated, as the parent proliferating cells, located in this outer shell, placed their daughter cells randomly upon division. 
On the contrary, dying cells in the necrotic core left empty spaces. 
This, in turn, led to a depression that was counterbalanced by the cells in the outer shell, resulting in an advective motion. 
The authors introduced novel metric parameters (such as the waiting time of cells in the proliferating rim or the radial infiltration velocity in the shell between the rim and the necrotic core) whose values were mapped to the composition of the simulated spheroids. 
This, in turn, provided a new way to infer the underlying morphology (e.g. the quiescent area) from measurements of the microbead's trajectories. 
Another approach to modelling tumour cells’ flows was presented by Jamous et al.~\cite{Jamous_et_al:2020}, which simulated oncostreams (i.e.~cells migration in opposite directions) and flocks (i.e.~cells migration in the same direction) in 2D and 3D. 
They reported that the presence of oncostreams correlates with tumour progression, while they also interrogated \insilico{} the parameter space impacting the mode of cancer cell migration. 
The authors showed that the formation of flocks in 2D simulations augmented as the cells’ shape was shifted from round to ellipsoid. 
Using the total polarization of the configuration as a proxy, the higher steering capability of the cells (which correlated with the eccentricity of their shape) was found to be the reason behind the increased flock formation. 
As the simulation domain was extended from 2D into 3D, cells were provided with an additional degree of freedom. 
While both streams and flocks emerged at low cell numbers, only the streams were observed at higher densities. 
Moreover, the authors found that the tumour dynamics is strongly affected by the cell density and that both flocks and streams emerge when the ability of the cells to steer drops; thus, dismantling of oncostreams was proposed in \cite{Jamous_et_al:2020} as a new therapeutic approach. 
Another agent-based model of tumour cell movement was presented in the work of Suveges and colleagues \cite{Suveges_et_al:2021}, who put emphasis on the role of the extracellular matrix (ECM). 
More specifically, the authors developed a hybrid multi-scale model to investigate if and how the ECM could impact the cell invasion patterns of cancer cells. 
Cells were modelled using an agent-based model which was linked to, affected and was affected by a continuous model of the ECM. 
To simulate the adhesive interactions between the cells and the ECM (that don't occur at a single point of contact), the authors employed non-local adhesion integrals.
These allowed long-distance interactions to be taken into account by defining a sensing region over which the adhesion strengths were integrated.
Their model demonstrated that aligned ECM fibres are necessary for tumour aggregations to move, while tumour invasion is impaired when the matrix fibres are aligned in parallel to the tumour margin.
Importantly, their \insilico{} findings were confirmed with experimental results, and they argued that the tumour expansion speed could be predicted from the orientation of the ECM fibres.

The impact of neoplastic cell heterogeneity on the tumour shape was further investigated in \cite{Gong_et_al:2021}, where a spatially resolved agent-based model was combined with a quantitative systems pharmacology model (QSP). 
The QSP, an immuno-oncology mathematical model of ordinary differential equations (ODEs), was used to simulate interactions among multiple compartments at the tissue scale. 
Interestingly, the propagating front of the tumour-enhancing region and the tumour necrotic core were simulated by two different agent-based models, while the effect of immune checkpoint inhibitors (such as anti-PD-1) on the tumour growth was also simulated. 
Virtual patients and clinical trials were probed using sets of parameters generated via Latin Hypercube sampling, and their agent-based model aided in identifying predictive biomarkers for the tumour diameter, anti-PD-1 responsiveness and time to cancer progression. 
Spatial heterogeneities within the tumour volume are not limited to cell morphologies, but rather encompass substance concentrations (e.g., glucose and oxygen). 
Modelling of these transients is well suited to agent-based models and allows, among other uses, inspecting the effect of local concentration changes on the cell cycle. 
Representative works that focused on the latter aspect were those of Hong et al.~\cite{Hong_et_al:2018} and Kempf et al.~\cite{Kempf_et_al:2015}. 
These models provided insights into the role of hypoxic conditions in cancer treatments, assessing the capability of hypoxia-activated pro-drugs in killing heterogeneous bystander cells otherwise unreachable.
Moreover, they highlighted the importance of timing and hypoxic sensitisers to maximise the efficacy of radiotherapy.

\subsubsection{Somatic cells' role}

Inherent components of the environment that surrounds the tumour tissue, somatic cells can contribute to neoplasia and promote its development. 
Amongst them, immune cells and, in particular, lymphocytes, play key roles in the process and thus provide an interface for testing new treatment modalities. 
Contextually, this section summarizes some of the latest attempts at modelling the interplay between somatic and cancer cells, while potential therapeutic approaches are outlined below.

In the work of Gong et al.~\cite{Gong_et_al:2017}, they emphasised the spatial patterns of ligand PDL1 that is secreted by immune cells as a way to inhibit excessive activity but also by cancer cells after protracted exposure. 
The high spatial resolution of the agent-based model developed provided the \insilico{} framework to correlate numerically pre-treatment immune architecture, patients’ features and immune checkpoint inhibitor outcomes. 
Moreover, they attempted to predict treatment responders using a threshold on the distance between the PDL1-positive cells and the tumour surface. 
The model was further extended by the authors in \cite{Gong_et_al:2021} (see previous paragraph) and \cite{Ruiz-Martinez_et_al:2022} by introducing a QSP model. 
Notably, the QSP module was used to simulate the human body in a 4-compartment model, where the bloodstream served as a source for T cells and myeloid-derived suppressor cells.
While the QSP employed ODEs to simulate the dynamics of the whole tumour at the tissue scale, the agent-based model replicated local changes at the cell scale in a small representative region. However, an ODE version of the agent-based model was built to keep consistency between the agent-based and the QSP models and the two were sequentially solved and used to update each other, with the input values scaled/inversely scaled properly. The effects of immunotherapy (i.e., anti-PD1) and different values of the migration and proliferation rates of the cells (encoded by adimensional parameters) on the tumour morphology were investigated. Additionally, an innovative use of a 2D Gaussian kernel density to smooth the discrete spatial distribution of the cells allowed the authors to introduce a new way to locate the boundaries of the tumour-invasive front from digital pathology images. A different therapeutic approach, namely chimeric antigen receptor (CAR) T-cell therapy, was modelled and investigated in \cite{Prybutok_et_al:2022}. 
Simulations of both a dish and a tissue (where nutrients are thus provided by the vasculature) resulted in the identification of the best treatment strategy, maximising cancer cell death by CAR T-cells while minimising the elimination of low-level antigen-expressing healthy cells. 
The work of \cite{Beck_et_al:2019} and \cite{Khazen_et_al:2019}, provide tools to characterise the immune response mediated by cytotoxic T lymphocytes (CTLs). In the first one, a simple model with space competition was used to prove that both CTL contact and cytokine secretion are needed for tumour cell killing. In the second, where CTLs could adhere to tumour cells, the optimal effector/target ratios for tumour control were found and the human CTL killing per capita was quantified. Under certain circumstances, immunotherapy alone might not be enough to eradicate solid tumours. The study of \cite{Kather_et_al:2017} reports that for patients affected by microsatellite-stable colorectal cancer, effective immunotherapy strategies don't exist. However, the agent-based model implemented by the authors shows that a combined therapy aimed at boosting the immune system while targeting the stroma can eradicate the simulated tumours in 75\% of the runs. In fact, a permeable stroma allows the lymphocytes, whose number is elevated following an external injection, to effectively counteract the immune evasion of the cancer cells and avoid the inhibition of the cell-killing mechanisms. Finally, viral-infected cells are used as a proxy to activate the CD4+ T cells in the model outlined in \cite{Jenner_et_al:2022}, where the authors examined the impact of the relative density of the stromal cells on the efficacy of oncolytic viruses (OV) for GBM treatment. The OV, simulated as a diffusing field, are uptaken by both cancer and stromal cells and the intracellular dynamics is modelled with ODEs. However, while OV replication and subsequent lysis occur in cancer cells, the stromal ones act as sinks and reduce the overall viral infiltration. As a consequence, cytokines are not released and the response of CD8+ T cells is hampered. The simulations, whose outcomes were validated against heterogeneous patient samples, showed that high viral biding rates could be ineffective if the relative density of GBM cells is low and only an increase in the number of CD8+ T cells led to a significant reduction of the tumour size.

The interaction between myeloma cells and bone marrow stromal cells was investigated in \cite{Su_et_al:2014} and \cite{Ji_et_al:2017}. Myeloma cells are thought to closely collaborate with bone marrow stromal cells in a positive-feedback loop that leads to niche stiffening and mechanical protection from drugs. As multiple myeloma has proven to be able to develop multi-drug resistance and evade the host immune response and relapse, combining multiple therapies could lead to improved outcomes.
Both the aforementioned models are robust \insilico{} procedures to test the joint efficacy of different anti-cancer drugs. To quantitatively measure the synergistic effects of the drugs, the authors employed the Loewe combination index \cite{Zhao_et_al:2004}. Moreover, both the models are hybrid and multi-scale in that ODEs are used to simulate intracellular dynamics, while agents simulate the cells. Besides, Ji et al.~\cite{Ji_et_al:2017} built upon the work of Su et al.~\cite{Su_et_al:2014} by implementing an immune system within the model. 
The authors simulated drugs that could target: the myeloma cells, their immune tolerance, the biomechanical phenotype of the bone marrow stromal cells and the communication between the latter and the myeloma cells. The models, which successfully replicated the tumour growth and interactions with the host cells, provide valuable resources to determine the efficacy of multi-drug treatments and the most promising dose combinations. 

Other noteworthy models of interactions between cancer and blood cells, as well as other host cells, were presented in the papers of Uppal et al.~\cite{Uppal_et_al:2014} and Heidary et al.~\cite{Heidary_et_al:2020}, where the role of platelets, key players in metastasis, and fibroblasts, turned into cancer-associated cells, was explored. 
In \cite{Kareva_et_al:2016}, ABM was employed to compare physiological wound healing and tumour-induced angiogenesis to interrogate the interplay between cancer cells and platelets. 
The model suggests that, by disrupting the physiological setting, tumour edges interfere with the well-orchestrated release of angiogenesis inhibitors, resulting in a `\emph{wound that never heals}' condition.
The role of the cross-talk between the endothelium and cancer was further explored by Yan et al.~\cite{Yan_et_al:2017}, who presented a hybrid model of GBM progression. 
Of note, the model features both normal endothelial cells and trans-differentiated vascular endothelial cells, together with neoplastic stem and differentiated cells. 
The \insilico{} results illustrated that the combination of therapies traditionally used in isolation can lead to enhanced results in GBM treatment. 
Modelling of the endothelium and the perivascular niche of the GBM was also explored by Randles et al.~\cite{Randles_et_al:2021}, where the authors employed ABM to optimise an existing therapeutic regimen.
Scalable simulations were combined with simulation annealing to infer the best timing for both chemo- and radiotherapy administration. 
The parameters obtained were then used to implement and test the schedule \invivo{}, which results in an improved outcome and thus provides experimental evidence for the initial assumptions regarding the stem-like cancer cell differentiation and translocation mechanisms.

\subsection{Machine learning approaches and ABM}
\label{sec:MachineLearning_ABM}

Typically formulated as an optimization problem, simulation calibration has become a very interesting candidate in the machine learning (ML) community. 
This is because, on one side, \insilico{} models can generate large quantities of data and, on the other side, biological or/and medical data is sometimes hard to collect or very expensive to acquire. 
Learning-based approaches offer an attractive alternative to optimization-based calibration approaches. 
They are especially interesting, as they need to also update behaviour rules embedded in ABM, as described in Figure~\ref{fig:1}b. 
Such approaches tackle the realistic reproduction of mechanistic dynamics in biological systems by learning the mapping from clinical data and model parameters to a performance metric. 
In this space, the approaches are rather diverse. For instance, the work of Barde et al.~\cite{barde2015direct} presented the first empirical application example of a novel probabilistic model calibration methodology. 
The system was designed to provide an information criterion on a given set of data for any model that is reducible to a Markov process. 
The rationale behind the development of this methodology was to allow the explanatory power of simulation models to be compared to more traditional modelling approaches suitable for clinical application. 
This has been identified as one of the main hurdles to the development of simulation methods, particularly in Monte Carlo-based models. 
Thus, in \cite{barde2015direct}, the end goal was to establish both the robustness of their respective calibrations and their explanatory power on the data.
Taking agent-based models closer to the data remains an open challenge, especially when considering biological processes. 
This aspect was the direct focus of the research carried out in the work of Lamperti et al.~\cite{lamperti2018agent}. 
The authors explicitly tackled parameter space exploration and calibration of agent-based models combining supervised ML. 
Together with intelligent sampling, the researchers proposed to build a surrogate meta-model; 
the meta-model provided a fast and accurate approximation of simulated model behaviours, dramatically reducing computation time. 
More precisely, the ML surrogate (i.e., an adaptive twin) facilitated large-scale explorations of the parameter-space, while providing a powerful filter to gain insights into the complex functioning of agent-based models capturing complex dynamics across scales. 
Using computational intelligence for learning ABM simulation parameters, Singh et al.~\cite{singh2018micro} employed artificial life optimization. 
Their \insilico{} framework implemented a hybrid model using micro-simulation and ABM techniques to generate an artificial society. 
The agents in this model derive their decisions and behaviours from real data (i.e.\ a micro-simulation feature) and interact among themselves (i.e.\ an ABM feature) to proceed in the simulation realization. 
Such approaches have been reported to map very well on the problem structure, as it is typically found in cancer cell biology, where local cell behaviours propagate in upper tissues or organ properties change.

The work of Niida et al.~\cite{niidaetal2019sensitivity} proposed a very computationally powerful and parallelised approach to deal with model uncertainty (and its impact on calibration) in ABM. 
They highlighted the role of interactive visualisation to help identify suitable model parameters. 
This is crucial in handling the highly nonlinear dynamics of cellular interactions and cancer evolution. 
In this context, the calibration process of ABM simulations is structured around the concept of adaptability.
By adaptability, we refer to the fact that model parameters can impact one another, either via direct relationships (e.g., diffusion constant and physical properties of the extracellular matrix) or implications on system dynamics (e.g., vascularisation and cellular metabolism can both affect the growth dynamics of cancer). More precisely, as the model complexity increases, so do the constraints among the model parameters and their inter-dependencies with regard to a given summary statistics.

Using empirical priors and statistical learning, Lima and his colleagues \cite{Lima_et_al:2021} proposed a moment-based Bayesian inference to account for the stochasticity of the coarse-grained agent-based model in a tumour growth multi-scale model. 
The approach presented very clever methods for quantifying uncertainties due to limited temporal observational data of cancer growth and staging at different spatial and temporal levels. 
Overall, the approach reduced the computational time of ABM simulations while reliably/realistically capturing tumour dynamics and its inherent nonlinearity.
Using a hierarchical optimization simulation for calibrating the agent-based model, Amaran et al.~\cite{amaran2014simulation} integrated optimization techniques into simulation analysis. 
The primary goal of simulation-based optimization is to improve the performance of the models through Monte Carlo processes.
More specifically, the Monte Carlo simulation allowed the system to find the optimal set of parameters for a given criterion based on a modular thresholding method.
The work of Akasiadis et al.~\cite{akasiadis2022parallel} stands out through the ingenious use of a typical calibration methodology -- they employed the extreme-scale model exploration with the swift framework to simulate the growth of tumour spheroids and scaled their ABM simulations on a high-performance computing server.
They employed a scale simulator for tumour cell growth and a genetic algorithm as a heuristic search method for finding good parameter configurations in a feasible time. 
Their \insilico{} method only considered numerical optimization and the goodness-of-fit only captured the quantitative aspects of the calibration.
Finally, neural learning techniques were employed for comparing spatial simulations to tumour imagery, going beyond basic metrics retrieved from tumour images and ABM simulations. 
In this manner, such algorithms may evaluate the model fit quantitatively.
More recently, the work of Cess and Finley \cite{Cess_Finley:2023} employed representation learning and a neural network to project an input into low-dimensional space, which is a representative example. 
The authors utilise a neural network to represent the ABM simulations and tumour images as low-dimensional points, with the distance between them, serving as a quantitative indicator of their differences.

\subsection{Multi-scale or multi-level numerical methodologies in ABM}
\label{sec:MultiScaleLevel_ABM}

Multi-scale or, as some investigators prefer to use as a term, multi-level \insilico{} modelling systems in biology and pharmacology are robust methodologies that can aid researchers in understanding and probing the fundamental mechanisms of biological phenomena and in clinical applications.
The excellent review of Morvan \cite{Morvan:2013} lays out clearly the definition of multi-scale methodologies using agent-based models, despite in his survey he adopts the term `multi-level agent-based modelling' instead of `multi-scale' since the latter, according to him and to Gil-Quijano et al.~(cited therein), has a restrictive meaning as it focuses on the spatiotemporal extents of levels and not on the interactions and organization within the biological system under investigation.
We will thus use the term of a `multi-level agent-based model' in this survey whereas, for semantic reasons, we will preserve the term `multi-scale model' if used in the cited works below.
Some examples of early attempts at multi-level approaches using agent-based modelling that considered coupling cell-scale to molecular-scale dynamics include the paper of Athale et al.~\cite{Athale_et_al:2005} who presented a model of gene–protein interactions integrated in an agent-based model system to probe the ability of brain cancer cells to `switch' between migrating and proliferating phenotypes, while test how molecular species interact with other molecules within and across sub-cellular compartments. 
However, in order to keep the review relatively short, we limit the depth of our survey to the relevant multi-scale / multi-level ABM papers in biology and biomedicine of cancer presented during the last fifteen years. 
Older review articles, such as those of Deisboeck \cite{Deisboeck_et_al:2011}, Stamatakos \cite{Stamatakos_et_al:2013}, Walpole \cite{Walpole_et_al:2013} and their colleagues, present the highlights of published multi-scale \insilico{} models of biological systems in cancer, cardiovascular and biomedicine, and demonstrate those early successes using agent-based models in the respective context. 

There is a fair list of published papers that demonstrate the coupling between discrete systems -- as in the case of agent-based models -- and continuum-based or network models to describe the biological cross-talk amongst different spatial scales by considering pertinent numerical algorithms and techniques suitable for the modelling task.
Not directly linked to the dynamics of carcinogenesis, however, Montagna and her colleagues \cite{Montagna_et_al:2010} proposed an interesting multi-level ABM approach to simulate the dynamics of drosophila melanogaster morphogenesis. 
Drosophila embryo cells are modelled as agents that divide, move, secrete/uptake substances, while Montagna's method encompasses also the balance of the molecules that mediate cell-to-cell communication and a gene regulatory network to simulate the molecular biology of the cells, i.e., the reactions taking place inside the cells. 
Their multi-level ABM methodology was implemented via a multi-threaded discrete event scheduler using software \emph{Repast Simphony} to simulate the expression patterns of the embryo cells against experimental evidence from the \emph{FlyEx} database.
Zhang et al.~\cite{Zhang_et_al:2011} proposed an agent-based brain tumour model that encompasses intercellular level to describe cell--cell interactions intracellular-scale dynamics by employing a system of ordinary differential equations to describe selected molecular pathways relevant to glioblastoma multiforme pathophysiology (i.e., phenotypic switches in cells from migration to proliferation), and the `tissue-scale' to model the balance of chemo-attractants concentration (through isotropic diffusion, secretion and consumption by the cells).
The main focus of this work was in the design and development of their in-house {C++}/{CUDA} implementation of the multi-scale agent-based model, which was parallelized with respect to both the chemo-attractant diffusion and the intracellular signalling processes using graphics processing units (GPUs) computing.
They reported a considerable computational speed-up of the GPU-based design of the multi-scale ABM simulator compared to the one of a sequential design -- this was amongst the few early works that demonstrated the potential of multi-scale ABM to simulate real-time cancer progression.
Cai et al.~\cite{Cai_et_al:2013} developed a three-dimensional hybrid cellular automata model, which is part of the family of agent-based models, to study the dynamics of tumour spheroids and probe the effect of hypoxia, cell phenotypic behaviour due to microenvironment biochemical factors. 
They solved three coupled reaction-diffusion equations to simulate the dynamics of the ECM, oxygen and ECM-degrading enzymes and communicated in a partitioned fashion the solution to an ABM simulator of the cell dynamics similar to the methodological approach of Zhang.
Alfonso and his colleagues \cite{Alfonso_et_al:2016} presented a comprehensive cancer model to study \insilico{} immune cell infiltration and interactions in the breast ductal lobular epithelium.
Following an ABM formulation, their multi-scale model accounts for myoepithelial, luminal and immune cells (each type allowed to reside in a separate lattice in 2D space) whose behaviour included immune cell trafficking, cell migration, immunosuppression, epithelial cell proliferation, damage, programmed and induced cell death, and cell lysis.
Also, they modelled the transport and secretion of chemokines that control the induction of an immune response in the terminal ductal lobular units of the breast epithelium.
They calibrated their model from imaging data of immuno-histochemical epithelial, vascular and immune cell markers from healthy women, and they investigated recurrent inflammation during physiological menstrual cycles and normal hormone levels, while they analyzed \insilico{} parameter perturbations that can lead to carcinogenesis.
Interestingly in the latter part, as suggested by the agent-based model simulation results, they observed that epithelial damage induced higher variations in immune cell infiltration.
Later, Gong et al.~\cite{Gong_et_al:2017} presented a multi-compartmental, multi-scale model of tumour development and anti-tumour immune response, which included interleukin-2 (a cytokine attributed to immunological homeostasis and classification), cytotoxic T lymphocytes and neoplastic cells.
Cells were set to interact in an off-lattice 3D space and follow a set of rules including division, migration, cytotoxic killing and immune evasion.
They developed an in-house {C++} code for their ABM simulator to explore spatio-temporal tumour immune response to PD1 and PDL1 inhibition and employed Latin hypercube sampling for the sensitivity analysis of their agent-based model to generate parameter value combinations that can produce model prediction accuracy for a small number of samples.
Their \insilico{} results generated interesting findings on the spatial patterns of different cell types without treatment that resembled patterns reported in cancer patient biopsies, and that ABM-simulated response to anti-PDL1 treatment is affected by the neoantigen characteristics of a patient.
Letort and her colleagues \cite{Letort_et_al:2018} presented an open-source simulator, \emph{PhysiBoSS}, which combined intracellular signalling using Boolean modelling and multi-cellular dynamics and behaviour using ABM.
As a use-case to demonstrate their modelling tool was modelling cell-fate decisions in response to treatment of cytokine TNF in order to illustrate the cell--cell communications.
They also explored \insilico{} the effect of different treatments and the behaviour of several resistant mutants, while also testing the dynamics of cancer cell population with respect to the spatial heterogeneity of biochemical cues and resources, i.e.\ oxygen.
Pally et al.~\cite{Pally_et_al:2019} presented in a paper both their experimental and computational work to interrogate cancer cell migration into cellular interactions with the basement membrane (BM) and its remodelling, the transition from BM to type-I collagen, and the subsequent remodelling of, and migration within, type-I collagen in the context of early breast carcinomatosis.
They built a multi-scale 3D organo- and pathotypic experimental assay, with the ABM implementation based on a cellular Potts model using the open-source software \emph{CompuCell3D}.
The model encompassed cancer cell proliferation and apoptosis, cell adhesion with the ECM, and the TME with respect to matrix remodelling through reaction–diffusion–based morphogen dynamics of metalloproteinases (MMPs) and tissue inhibitors of MMPs.
Pally designed a culture model of MDA-MB-231 cells to form reconstituted BM-coated suspended clusters to mimic the invasion patterns of breast cancer cells \invivo{} and probe how ECM density, metalloproteinase and N-linked glycosylation concentration impacts cancer cell invasiveness.
Sfakianakis and his colleagues \cite{Sfakianakis_et_al:2020} presented a multi-scale modelling framework for cancer invasion of the ECM by the combined action of epithelial-like cancer and mesenchymal-like cells (ECCs and MSCs respectively).
Their approach considers a hybrid system of partial differential equations, for the spatio-temporal evolution of the densities of the ECCs, ECM and the MMPs, and stochastic differential equations, for the time evolution of the MSCs including their migration along ECM gradients with these cells described as particles.
The implicit/explicit Runge-Kutta finite volume numerical method was employed to solve the continuum-based part of the model, and an explicit Euler-Maruyama scheme was employed for the solution of the stochastic particle-based part of the model.
In the same year, Macnamara and her colleagues \cite{Macnamara_et_al:2020} proposed a multi-level \insilico{} method to simulate the growth of a solid tumour, migration of cancer cells within heterogeneous tissue, and the effects of fibre and vascular structure in cancer development.
In contrast to previous works, they modelled each cell as a viscoelastic deformable sphere with cell dynamics governed by ordinary differential equations to describe cell--cell forces, cell random migration and cell to ECM friction, and cell behaviour (i.e., cell cycle, growth and birth) as in agent-based models.
Cancer cells were allowed to occupy voxels within a Cartesian grid, which was also populated with additional agents that were contiguously structured such that they formed capillary vessels. 
They also accounted for the balance of nutrients and oxygen at the tissue level in the form of reaction-diffusion equations that were discretized with finite elements.
Macnamara built an in-house {C++} solver for cell dynamics and used the \emph{FreeFem++} platform as a finite element solver, while they ran (non-cancer specific) simulations of tumour growth around an arbitrary blood-vessel network.

De Montigny et al.~\cite{deMontigny_et_al:2021} proposed a hybrid multi-level \insilico{} cancer model that was tailored to simulate glioma growth.
Their off-lattice agent-based model encompasses (host and neoplastic) cell growth, division, migration and adhesion, the dynamics of the extracellular matrix, the effects of oxygen and nutrient availability in cell survival, or the switch of cancer cells into a hypoxic or necrotic state, and the signalling triggered by chemical cues and growth factors.
In contrast to all the above-mentioned papers, the multi-level formulation in \cite{deMontigny_et_al:2021} uniquely couples a continuum-based finite element model for the solution of reaction-diffusion equations (i.e., to predict the balance of cytokines, growth factors and oxygen) at the tissue scale with an agent-based model via the volume averaging method.
The hybrid simulator was developed by coupling an in-house FEM solver \emph{FEB3} and the open-source platform \emph{BioDynaMo} for the ABM simulations.
Using the \insilico{} framework, they examined the impact of cell--cell and cell--ECM interactions in (macroscopic) tumour growth, brain tissue perfusion and tumour necrosis, as well as they assessed computationally the differences between low- and high-grade glioma growth, vascularization and necrosis and compared to experimental data from the literature.

Lima et al.~\cite{Lima_et_al:2021} presented a reduced-order ABM methodology coupled with Bayesian inference modelling for parameter calibration to manage the stochasticity of the agent-based model.
In terms of the multi-scale modelling, ABM was used for the phenotypic behaviour and the geometric properties description of the cells, while the dynamics of nutrients was modelled at the tissue scale as a reaction--diffusion process.
Their hybrid model was demonstrated to simulate the development of a {BT-474} human breast cancer cells \invitro{}, using time-resolved microscopy data, and employed a moment-based Bayesian inference method to quantify the uncertainties owing to limited temporal observational data of carcinoma's growth.

Ponce-de-Leon and colleagues \cite{PonceDeLeon_et_al:2022} presented a multi-scale model of cancer cell dynamics with signalling for TNF-receptor dynamics, as in their previous intracellular signalling work in Boolean modelling \cite{Letort_et_al:2018}.
Cancer cells were modelled as agents residing in 2D or 3D lattice, that accounts for the presence of oxygen and the cytokine TNF.
Their \insilico{} model was built by combining the open-source ABM simulator \emph{PhysiCell} and the software \emph{PhysiBoSS} that was developed by the authors.
Subsequently, they integrated their simulator with an Extreme-scale Model Exploration with the Swift platform to carry out exploration tests of the agent-based model parameter space which was ultimately deployed to optimize dosage-specific treatments for tumour regression.
They probed for the effect of the spatial distribution of cancer cells on the treatment parameters optimizing the supply strategies in cell monolayers and three-dimensional tumour spheroids; similarly, they interrogated the robustness of the effective treatments with respect to the cell population heterogeneity of the cancer cells.
Following the modelling work in \cite{PonceDeLeon_et_al:2022}, Ruscone et al.~\cite{Ruscone_et_al:2023}
proposed an enhanced multi-scale model to interrogate possible targets that can help block or suppress the invasive phenotypes of cancer cells.
More specifically, the improvements are focused at the intracellular scale where they incorporated mechanisms of epithelial-to-mesenchymal transition and cell metastasis. 
They used the \insilico{} model to test the role of tumour protein 63 and metalloproteinase MT1-MMP in tumour invasion, as well as that of the tyrosine kinase protein SRC in an epithelial monolayer, while they also tested possible drug candidates to block migration in the ECM of cells that have undergone epithelial-to-mesenchymal transition.
Tsingos and her colleagues \cite{Tsingos_et_al:2023} presented a spatially inhomogeneous cellular Potts model to simulate cell migration in a fibrous matrix.
To overcome the substrate homogeneity of the Potts model, they coupled it with a `background' bead-spring biomechanical model of the ECM where fibre networks were modelled using molecular dynamics. 
A unique feature of their multi-level approach was the incorporation of contractile pulling by the cells through discrete focal adhesion-like sites on the fibre network.
Despite their simulation experiments coming from the angle of morphogenesis and tissue healing, their \insilico{} method can be adapted to study cancer cell infiltration and invasiveness.
Also very recently, Miller et al.~\cite{Miller_et_al:2023} presented a multi-scale modelling approach to evaluate the effect of chemotherapeutics on patient tumours based on metabolomic analysis results of lung cancer biopsy data.
Despite their multi-scale model being based on previous work from the same group (see reference [32] therein), they integrated metabolomic analysis evidence (from patient tumours) and modelling.
Interestingly, they built a synthetic dataset using Monte-Carlo by resampling selected parameter values of the multi-scale model to simulate chemotherapy, while they considered seven evaluation metrics to quantify the tumour response; however, as the authors suggest, their results need further validation with metabolic evidence from different and larger patient datasets.

The above literature survey illustrates a remarkable development track of multi-scale and multi-level methodologies using ABM in oncology over the past decade.
The computational cost and complexity to test the numerical stability of a multi-scale / multi-level \insilico{} procedure, to probe the sensitivity of the numerical schemes involved, and to calibrate the models across multiple spatial scales remains a challenge however.
In addition to the computational cost emerging when it comes to simulating thousands or even millions of agents and the inherent stochasticity of an agent-based model, this stipulates realizing a great number of simulations to accurately represent the statistical features of an \insilico{} cancer model.
Thus, the effort towards achieving a high level of robustness and fidelity in a multi-scale / multi-level ABM formulation elevates tremendously in proportion to increasing the size of the biological system under investigation, as well as with the quantity and the modalities of data coming from the laboratory or the clinic.

Interestingly, as it is presented in subsection~\ref{sec:MachineLearning_ABM}, investigators have attempted to amalgamate sophisticated machine learning and optimization algorithms for learning the simulation parameters, to quantify the model uncertainty and its impact on calibration on agent-based models (e.g., as in \cite{singh2018micro} and \cite{Lima_et_al:2021}). 
However, the majority of multi-scale and multi-level methodologies have used data to constraint and validate the cancer model predictions on a single scale, e.g., usually at the tissue level through tumour size measurements, or in the order of the largest temporal scale, e.g., typically at the order of the time-duration of a preclinical cancer experiment.
The cited papers that follow illustrate the gradual evolution of relevant cancer \insilico{} models to accomplish calibration across the scales.
Among the early attempts to simulate multicellular tumour spheroids that mimic the TME dynamics was that of Cai et al.~\cite{Cai_et_al:2013}.
Their multi-level approach -- cells were modelled as agents while the balance of nutrients and enzymes was modelled following a continuum-based method -- the simulation predictions were tested using history plots of the tumour spheroids with respect to size.
Later, following a similar modelling path, Mao et al.~\cite{Mao_et_al:2018} presented a hybrid continuum/agent-based model for HCT116 tumour spheroids to simulate hypoxia-dependent interactions between ionising radiation and a hypoxia-activated prodrug; 
this \insilico{} tool was used by Hong et al.~\cite{Hong_et_al:2018} to build a PK/PD model and probe the bystander effects of hypoxia-activated prodrugs in cancer cell killing.
To inform their model at the continuum scale (i.e. the average rate of diffusion of the drug molecules in the medium) they adopted parameters from the literature, while to calibrate the agent-based model they used \invitro{} data coming from flow cytometry analysis, confocal microscopy image data of the spheroids, and fluorescent staining of the cancer cells to mark their protein expression.
In \cite{Kather_et_al:2017}, they presented an \insilico{} model of lymphocyte–tumour–stroma interactions to interrogate the response to immunotherapy and stroma-targeting therapies on human colorectal cancers.
As in \cite{Hong_et_al:2018}, Kather et al.~employed data from horizontal \invitro() migration experiments on lymphocytes to inform the agent-based model, as well as \exvivo{} measurements (based on morphological processing of single slice images) on histological human tumour tissue samples to estimate the proliferation, apoptosis and distance to necrosis parameter values, and provide a quantitative basis for the cell-scale modelling.
Rahman et al.~\cite{Rahman_et_al:2017} coupled in `space and time' multi-scale cancer model spanned from tissue (using FEM) to cellular (using ABM) and subcellular scale, with the latter being represented by signalling pathways.
In a similar fashion, de Montigny et al.~\cite{deMontigny_et_al:2021} integrated ABM with the FEM using a volume-averaging formulation to build a multi-level brain cancer simulator.
Both models however were calibrated and tested from observed data at a single scale (tissue level), i.e., history plots of the tumour volume and average volume fraction of cell groups.
With an exception to the model in \cite{deMontigny_et_al:2021} where tissue-scale (FEM) parameters for tumour and host cell dynamics were inferred (a process often called data upscaling) from the cell-scale (ABM) simulation predictions, in \cite{Rahman_et_al:2017} the modelling parameters on the lower scales (cellular and subcellular) were either adapted from the literature or estimated.
The multi-scale approach of Lima \cite{Lima_et_al:2021} employed \invitro{} data to separately inform (at the cell scale) the mechanistic agent-based model of human breast cancer cells' phenotypic behaviour, and (at the continuum tissue scale) the transport and balance of glucose concentrations and cytokines in general.
The important contribution of their paper concerns the Bayesian inference concept applied for the time-dependent sensitivity analysis of the \insilico{} model and to interrogate the model parameters' space. Using lightweight neural networks Axenie et al.~\cite{Axenie_Kurz:2020} extracted the mechanistic relations governing phenotypic staging and tumour volume development \cite{axenie2021glueck}.
Very recently Cesaro and her colleagues \cite{Cesaro_et_al:2022} demonstrated their multi-level TME simulator that couples mechanistic agent-based models with PDE-based solvers in two dimensions.
An innovative feature of their paper was the data-driven strategy they adopted to inform the agent-based model using bulk gene expression data from The Cancer Genome Atlas database.
They also used evidence from the scRNA-seq dataset of human colorectal cancers to calibrate with respect to the tumour mutational burden and the inhibitory immune checkpoint that suppresses T-cell activation and to estimate the cell (HCA, T-reg, CAF, immune) fraction and in their model.
The multi-scale model of Ponce-de-Leon \cite{PonceDeLeon_et_al:2022} considered the Covariance Matrix Adaptation Evolutionary Strategy for the numerical optimization of their agent-based model to analyze the treatment parameters of the tumour necrosis factor cytokine and its effect on cancer regression.
Despite their model being inherently multi-scale, their \insilico{} experiments were calibrated on macroscopic quantities, i.e., drug dose, injection time and duration, and tumour size.
In the same year however, Miller et al.~\cite{Miller_et_al:2022} presented a multi-level approach that is unique in that it proposes to link tumour metabolomic measurements from patients into the mathematical model for tissue-scale behaviour of a carcinoma progression or control, the development of angiogenesis, the effect of chemotherapy, etc.
An important limitation however of the cancer model concerns that its behaviour depends mainly on the metabolomic data available, and how they are appropriately weighted and combined to determine the effect on the (mechanistic) model parameters.
Finally, Ruiz-Martinez and colleagues \cite{Ruiz-Martinez_et_al:2022} proposed a hybrid ODE-based / ABM simulation tool to investigate therapeutic strategies related to anti-cancer immunity and immune checkpoint inhibition.
The rules for the cancer (stem-like, progenitor and senescent) and the immune (CD8+ T and Treg) cells were defined in the agent-based model provided \invivo{} evidence from the literature \cite{Norton_et_al:2018}, while for the calibration of the differential and algebraic equations (120 in total) they employed a QSP model.
Their QSP, a mechanistic modelling method often used for drug discovery, was based on a relevant one for non-small cell lung cancer and incorporated data from single-cell RNA sequencing.

\subsection{ML-assisted calibration: a two-sided sword}
\label{sec:ML_assisted_Calibration}

Agent-based numerical procedures that are supported by ML models, as has been briefly outlined in subsection \ref{sec:MachineLearning_ABM}, can significantly contribute to the generation of agent-based models with suitable model parameters. 
However, automated calibration can also entail challenges and difficulties. 
This section elaborates on the comparison between ABM with and without ML assistance. 
Notably, we provide further explanatory comments to establish the wider context of ML and its involvement in ABM.

From a modeller's perspective, a pure ABM procedure is, as is more generally mechanistic modelling, often seen as demanding with regard to the determination of model variables. 
As explained above, ML techniques can be employed to \textbf{efficiently search the parameter space} of mechanistic models and determine optimised model parameters. 
However, an important criterion in biomedical models is that parameters need to be `\emph{biologically plausible},' if not (ideally) directly experimentally informed model parameters. 
An agent-based model where model parameters were estimated from the literature, without ML-assisted optimisation, is the study of \cite{Macklin_et_al:2012}. 
However, due to limitations in data availability, it is usually impossible to infer all model parameters from the literature. 
ML can help address this problem. 
For instance, Demetriades and his colleagues \cite{Demetriades_et_al:2022} employed ML to infer various parameters on the pharmacological impact of cancer drugs. 
B{\"o}rlin and his colleagues \citep{Borlin_et_al:2014} employed model parameters obtained both from the literature as well as ML-derived ones. 
\cite{jalalimaneshetal2017simulation} make use of reinforcement learning to optimise radiation treatment. 
Overall, the need to account for biomedical plausibility in model parametrization highlights the importance of interdisciplinary collaboration particularly for mechanistic, multi-scale models \citep{West_et_al:2022} (more about multi-scale models in subsection \ref{sec:MultiScaleLevel_ABM}).

A crucial criterion in computational modelling is `\emph{explainability}.' 
This aspect has recently gained much attention due to the fact that large-language models (LLMs) are very problematic with regard to gaining insights into the human-understandable causes of outputs. 
Especially when it comes to biomedical applications, explainability is of utmost importance. 
Clinicians need to understand exactly the reasons behind their findings to adequately inform their decision-making process, given the potentially life-altering impacts of these. 
Mechanistic models, as in for example the studies of Macklin \cite{Macklin_et_al:2012} and de Montigny \cite{deMontigny_et_al:2021}, produced experimentally verifiable hypotheses that can ultimately lead to deeper insights on glioma cancer growth and cell necrosis.
Notably, many ML methods suffer from the same issues as LLMs as they can be perceived as black boxes that may have excellent performance with regard to a given biomedical problem, but nevertheless limited clinical benefits. 
Therefore, modellers should be aware of the potential pitfalls when interfacing mechanistic models with ML algorithms, especially when it comes to practical impact in the clinical setting. 
However, it is notable here that ML methods come in different types, i.e., fundamentally black-box and white-box models. Hence, ML approaches that enable explainability (or interpretability) exist, as shown in the work of Linardatos et al.~\cite{linardatos2020explainable}. 

In view of the trend of \insilico{} models' rising complexity and the number of parameters it consists of, then comes the need to rigorously assess `\emph{robustness}' and `\emph{sensitivity}' of the model, for instance with regards to the model parameters and/or different initial conditions. 
Given that no two biological systems are exactly identical, a theoretical model must tolerate changes to model parameters, at least within reasonable boundaries. 
Due to this inherent variability, biological systems in cancer usually comprise redundancy, checkpoints and control loops that permit for changes without endangering important outcomes. 
In the context of computational modelling, limited changes to model parameters should not lead to implausible \insilico{} outcomes. 

Additionally, a crucial goal of computational modelling is the generation of hypotheses and experimentally verifiable predictions. 
To this end, ABM sensitivity analysis can help gain insights into the impact of the model parameters. 
A direct but basic way to accomplish such tests for `robustness' and parameter `sensitivity' is to execute the model with different model parameters in a grid-like manner -- in this methodology, ML cannot be deemed pertinent. 
However, it may come to a case where some parameters of an agent-based model may have a stronger impact than others. 
In this direction, ML can be a helpful tool to efficiently sample, identify and rank model parameters in terms of their importance when considering these as `features.' 
This is called Variable Importance Analysis (VIA) and, for instance, the random forest ML method is commonly used for such analysis, as demonstrated in the work of Pereda et al.~\cite{pereda2017brief}. 
In a similar approach, Retzlaff and his colleagues \cite{retzlaffetal2023integration} used decision trees for VIA in their agent-based model; 
the authors indicated that cell cycle duration and motility in the context of solid tumour metastasis are the most important factors with regard to therapy resistance. 

An aspect that requires consideration for the usage of mechanistic modelling and ML-assisted mechanistic modelling is `\emph{scalability}.' 
Given that mechanistic modelling should be, for the sake of biological plausibility, based on local information exchange only, its simulation can naturally make use of parallelised and distributed computing. 
However, adapting ML methods for large-scale applications requires often custom efforts, since every algorithm has a distinct communication pattern, as demonstrated in the work of Verbraeken et al.~\citep{Verbraeken_et_al:2020}. Along those lines, synchronisation requirements among nodes can vary across ML methods, as well as suitability for specific hardware (e.g., {CPUs} versus {GPUs}).
Of course, recent demonstrations of LLMs show that certain ML methods can be trained and employed in a highly performing manner \citep{narayananetal2021efficient, aminabadi2022deepspeed}. 
But the smooth and efficient interfacing with mechanistic modelling constitutes nevertheless a challenging task that remains to be addressed in the future.

Clearly, specific problems in cancer biomedicine require consideration of the associated advantages and disadvantages, and there is no one-size-fits-all approach for any given scientific quest. 
ML methods are widely and, due to their recent successful applications in many domains, increasingly used in combination with ABM and/or mechanistic modelling. 
However, a particularly intriguing research direction is the reverse direction: ABM can in principle be used to devise novel ML methods \citep{liuetal2021cooperative, gronauer2022multi}. It remains to be seen how such an approach can be leveraged for cancer biomedicine.

Overall, ML can be a highly valuable asset for researchers employing ABM. However, the combination of these distinct approaches can also entail challenges, and so their symbiotic application is not necessarily warranted. Nevertheless, specific aspects that need to be considered on a case-by-case basis can be appreciated (Table~\ref{tab:1}).

\begin{xltabular}{\textwidth}{|X|X|X|}
\caption{Comparison of pure mechanistic modelling (MM) versus ML-assisted mechanistic (ML-MM) approaches.}
\label{tab:1}
\\ \hline
\multicolumn{1}{|c|}{\textbf{Criterion}} & \multicolumn{1}{c|}{\textbf{Well-suited for}} & \multicolumn{1}{c|}{\textbf{References}}
\\ \hline 
\endfirsthead
\hline 
\multicolumn{3}{|r|}{{Continued on next page}}
\\ \hline
\endfoot
\hline
\endlastfoot
Identification of model parameters & ML-MM & \citep{Cess_Finley:2023,PonceDeLeon_et_al:2022,jalalimaneshetal2017simulation,Borlin_et_al:2014}
\\ \hline
Accordance with experimentally measured parameters & MM & \citep{cognoetal2022lungfibrosis,Macklin_et_al:2012}
\\ \hline
Explainability & MM & \citep{Macklin_et_al:2012,deMontigny_et_al:2021,PonceDeLeon_et_al:2022}
\\ \hline
Robustness and sensitivity & ML-MM & \citep{pereda2017brief,retzlaffetal2023integration}
\\ \hline
Computational scalability & MM & \citep{breitwieseretal2023highperformance,clascaetal2023lessons}
\\ \hline
\end{xltabular}

\subsection{Fusing mechanistic and learning approaches (physics-informed systems)}
\label{sec:Fusing_Models}

The \insilico{} modelling approaches discussed above focus on the calibration of Markov-chain / ODE- / PDE-based models to simulate processes involved in neoplasia. 
Such mechanistic models account for assumptions about the dynamics of the systems in both the temporal and spatial dimensions, as outlined in subsection~\ref{sec:MultiScaleLevel_ABM}. 
Advanced numerical methods and high-performance computing enable high-fidelity simulations of such calibrated mechanistic models to run at scale. 
However, most approaches for calibrating biological systems' models focus on fit quality. 
The common noun in the current approaches landscape demonstrates that the calibration error, which varies depending on the optimization approach, reaches an insurmountable barrier that can result in a standstill in selecting the ``optimal'' model.
This model selection process reaches another stale point when considering capturing corner cases in the spectrum of the modelled system's behaviours. The current approaches look at modifications of the models themselves with adaptive features or the analysis of the influence of the system's characteristics on the system's behaviour in corner-case situations.
Learning-based approaches offer an attractive alternative to these optimization-based calibration approaches. 
An additional advantage is the explanatory power that physics can offer when building ABM behaviour rules for the simulation.
Such approaches tackle the realistic reproduction of the biological system's behaviours by learning the underlying mapping from data and model parameters to a performance metric or goodness-of-fit criteria of plausibility.
But, in order to gain the best of the two worlds, mechanistic `biases' can be `injected' into ML models and leverage the power of learning from large amounts of data through a `directed' search for the solution, in other words, the realm of physics-informed ML modelling. 
Fusing mechanistic biophysical models and learning algorithms amounts to introducing appropriate observational, inductive or learning biases that can direct the learning process towards reaching physically plausible solutions.
This new conceptual framework of physics-informed ML framework coined by Karniadakis and his team \cite{karniadakis2021physics} proposed training ML models from additional information obtained by enforcing the physical laws (for example, at random points in the continuous space-time domain). 
Such physics-informed learning integrates (noisy) data and mathematical models, and implements them through neural networks or other kernel-based regression networks for calibrating or optimizing ABM parameters. 
Practically, this can be done by introducing inductive, observational, or learning biases in the learning process, under the form of a loss function, regularization term, or event calibration metric.
Multiple candidate approaches that focus on physics-informed learning for simulation calibration have been proposed, each one focusing on a different component of the overall problem. 
The incremental mixture approximate Bayesian computational procedure presented by Rutter et al.~\cite{rutter2019microsimulation} for colorectal cancer simulation calibration, using a simulated sample from the posterior distribution of model parameters given calibration targets in order to inform national cancer screening guidelines.
For instance, in order to achieve computational gains in large-scale simulations the work of Wood et al.~\cite{wood2008fast} developed a novel computationally efficient method for direct generalized additive model smoothness selection. 
Designed as a highly stable, but carefully structured, calibration system, the proposed approach achieved a computational efficiency that led, in simulations, to lower mean computation times than the schemes that are based on working model smoothness selection. 

Because it may be challenging to abstract and define the rules that control an agent-based model from experimental data, at least in an objective manner, there is a particularly synergistic potential to utilize ML to help infer the most effective, system-specific ABM rules, as shown in the work of Sivakumar et al.~\cite{sivakumar2022innovations}. 
Once such rule sets are developed, a large volume of ABM simulations can produce a plethora of data, and ML can be used in that setting as well. 
For instance, statistical measures that accurately and meaningfully characterize the stochastic outputs of a system and its features are one use of ML in this context. 
ABM simulations can produce credible (realistic) datasets to subsequently use for training ML algorithms (e.g., for regularization, to prevent overfitting), as an example of synergy in the other direction (from ABM to ML).
In an effort to develop a general-purpose computational framework, Spolaor et al.~\cite{spolaor2019coupling} introduced a novel approach for the analysis of hybrid models consisting of a quantitative (or mechanistic) module and a qualitative module that can reciprocally control each other's dynamic behaviour through a common interface. This qualifies as a mix of inductive and observational biases. More precisely, the system of Spolaor et al. took advantage of precise quantitative information about the temporal evolution of the modelled system through the definition and simulation of the mechanistic module. At the same time, it described the behaviour of biophysical model components and their interactions that are not known in full detail, by exploiting fuzzy logic in the definition of the qualitative module. 
Such approaches are deemed to be suitable for the analysis of cancer morphogenesis, an intricate chain of biological mechanisms that enable cell populations to reproducibly self-organize into specific shapes or patterns. 
Through physics-informed simulations, a modeller can modulate the state through controlled signal transduction on a range of spatial and temporal dimensions that include a variety of mechanisms and systems, as demonstrated in the work of Glen et al.~\cite{glen2019agent}.
The path of hybrid approaches is further strengthened by the work of Ward et al.~\cite{ward2016dynamic} which proposed a dynamic calibration of agent-based models using data assimilation. 
More precisely, investigators tackled the question of how such models can be dynamically calibrated using the ensemble Kalman filter, a standard method of data assimilation. 
The work developed a new type of Kalman filter-based system in a simple setting for data assimilation and fusion in ABM calibration for cancer development.
Combining probabilistic machine learning in a physics-augmented framework, the work of Moon et al.~\cite{moon2018data} claimed a new technique dedicated to improved calibration and validation of agent-based models. 
The framework identified periods of deviation between the simulation and the observation with the Hierarchical Dirichlet process hidden Markov model. 
This allows the framework to automatically calibrate the temporal macro parameters by searching parameter spaces with a broader likelihood of validation for tumour growth under a compromised immune system. 

When considering clinical sequencing of surgery and chemotherapy, the work of Axenie and Kurz~\cite{Axenie_Kurz:2020} illustrated how a physics-informed ML system can extract the pharmacokinetics of a common breast cancer chemotoxic medication while also concurrently learning the patterns of tumour development in a variety of breast cancer cell lines.
In a very recent study, Beik et al.~\cite{beik2023unified} introduced a Bayesian multi-model inference methodology for a dual purpose. 
On one end, the model quantified how mechanistic hypotheses can explain given experimental datasets, basically by attaching the probabilistic explanation to data peculiarities. 
On the other, the model demonstrated how each dataset informs a given model hypothesis, thus, enabling hypothesis space exploration in the context of available data. 
The approach was successfully used to probe standing questions about heterogeneity, lineage plasticity, and cell--cell interactions in tumour growth mechanisms of small-cell lung cancer. 
The methodological approach in \cite{beik2023unified} complements the physics-informed ML approaches with a strong probabilistic framing of hypothesis testing and variable interactions in cancer modelling. 

On the other end of the spectrum, when considering the translation of \insilico{} models in cancer progression, physics-informed ML approaches have been successful in tumour volume prediction after learning without supervision tumour phenotypic stages from breast cancer cell lines (e.g., \cite{Axenie_Kurz:2020}) as another simultaneous task. 
For the purpose of optimization-free calibration of ABM simulations, the work of Axenie et al.~\cite{axenie2022fuzzy} introduced a physics-informed fuzzy logic calibration system. 
Using spatiotemporal models of agents' interactions, the \insilico{} system could regress, based on human experts, the plausible solutions of the goodness-of-fit metric (e.g. Akaike Information Criterion (AIC), Root Mean Squared Error (RMSE) etc.).
Benefiting from expert knowledge, known physics models and inference capabilities, the calibration framework in \cite{axenie2022fuzzy} provided a very good trade-off between plausible/realistic reproduction of real dynamics, plausible choice of model parameters, and a very fast calibration procedure. 
From a methodology point of view, this approach is superior to those using solely optimization algorithms, as in \cite{akasiadis2022parallel} for instance. 
Akasiadis and his colleagues only considered numerical optimization, where the goodness-of-fit only captured the quantitative aspects of the tumour growth agent-based model calibration. 
Yet, when considering the plausibility of the candidate solution, the numerical approach might offer a, sometimes, non-intuitive or plausible parametrization of the agent-based model behaviour rules. 
This can be overcome by infusing a mathematical description of tumour growth covariates or other mechanistic dependencies in the learning function (i.e. modelling a loss function of the log-likelihood distance from the data to the mechanistic model).
The different approaches for physics-informed learning calibration of ABM simulations demonstrate the potential such an approach has to leverage known models and learning algorithms and demonstrate how their combination is beneficial to achieve plausible, realistic simulations. 
Thus, physics-informed machine learning offers an attractive numerical procedure for extracting an accurate human-understandable representation of the underlying dynamics of physical interactions crucial to typical oncology problems, as demonstrated by the very encouraging results from multiple predictive tasks instantiations in oncology, as shown in the work of Kurz et al.~\cite{kurz2021data}.
This overview highlights the way ML systems may enhance clinical decision-making using effective computational techniques that benefit from embedding priors in the learning processes in order to guide their convergence towards plausible solutions. In order to do this, we think that such platforms provide a link between the modeller, the data scientist, the data, and the practising physician.

\section{Strategies for agent-based models calibration and validation}

\subsection{Calibration as a multi-stage validation} 
\label{sec:Calibration}

Independent of the underlying parameter inference approach, a calibration procedure typically comprises multiple stages: 
data acquisition, scale choice (global or local), performance metrics ($M$) definition, the definition of the metrics for the goodness-of-fit ($G$), and the choice of an optimization algorithm. 
A first, and crucial question is what input data is available for an agent-based model calibration? 
\INVITRO{} cell line data is typically available but poised by small sample size, uneven sampling, and multi-modality. 
\INVIVO{} data is usually collected under strict protocols but typically covers narrow aspects of the study's phenomena. 
Second, the discrepancy in calibration is, typically, measured by the goodness-of-fit of the simulated parameters to the real parameters. This is supplemented by a series of calculated quantities by the metrics of performance in both the real world and simulation.
Every goodness of fit evaluation can be performed either globally (e.g., using Least-squared errors) or locally (e.g., using maximum likelihood). 
Since at its core, model calibration is basically an optimization problem, the underlying algorithm aims to converge to a solution that is close to the global minimum of the goodness-of-fit metric, $G$, while obeying imposed constraints on parameters' values. 
A formulation of the calibration procedure can be synthetically described as:

\begin{equation}
\begin{aligned}
\quad &\min_{M_{sim}, M_{real}}~f(M_{real}, M_{sim})
\\
\textrm{with}\quad &f(.) = G(.),~M_{sim} = F(\boldsymbol{x}|\boldsymbol{\beta}),~M_{real} = F(\boldsymbol{x})
\\
\textrm{s.t.} \quad & \boldsymbol{\beta_{min}} \leq \boldsymbol{\beta} \leq \boldsymbol{\beta_{max}}\,,
\end{aligned}
\label{eq1}
\end{equation}

\noindent
where $F$ is a mapping function of the biophysical system's temporal trajectory data $\boldsymbol{x}$ provided a set of model parameters $\boldsymbol{\beta}$ calculated after simulating the model to extract the simulated values of the performance metrics $M_{sim}$, with $M_{real}$ being the observed values of the performance metrics calculated by $F$ from data $\boldsymbol{x}$; 
$f$ is a function representing the goodness-of-fit $G$ (i.e., the realism of the simulation or closeness of $M_{real}$ to $M_{sim}$), and $f(M_{real}, M_{sim})$ is the objective function to be optimized calculated from $\boldsymbol{x}$ -- in principle a function describing the discrepancy between simulation and reality (i.e., $M_{sim} \approx M_{real}$).
A mapping of this formalism to the disease trajectory is given in Fig.~\ref{fig:2}b.
This formalism was adopted by many researchers in their attempts to calibrate ABM simulations.
We chose to analyse two very relevant candidates that currently capture the state-of-the-art strengths and limitations of such methods. 
Cess and Finley~\cite{Cess_Finley:2023} presented a novel approach that applied neural networks to represent both tumour images and ABM simulations as low dimensional points, with the distance between points acting as a quantitative measure of the difference between the two. 
This enabled the authors to extract a quantitative comparison of tumour images and ABM simulations, where the distance between simulated and experimental images can be minimized using standard parameter-fitting algorithms. 
We can see that each of the quantities in Equation~\ref{eq1} can take arbitrary dimensions, but there will always be a `distance-based' goodness-of-fit $G$ to characterise the plausibility of the simulation after calibrating the models' parameters $\beta$.
But, as the literature survey shows, there has so far been no shared view on the quantification of validity in agent-based simulations.
Very recently, Troost et al.~\cite{troost2023keep} conceptualized validation by systematically substantiating the premises on which conclusions from simulation analysis for a particular modelling context are built.
They provided a formal extension to the classical approach in Equation~\ref{eq1}. To this extent, a strict definition of the parameters $\beta$ is the problem dependent and the choice of functions $f$ depends on the trade-off between explanatory power, predictive accuracy, and the plausibility captured by the bounded $\beta$ values.
They proposed an assessment of the validity of agent-based models by incorporating valid conclusions from simulation analysis in a context-adequate method that touched model construction, model and parameter inference, uncertainty analysis, and the simulation process itself.

We now turn our attention to how optimization-based and optimization-free multi-stage methods complete the landscape of relevant approaches for multi-stage validation.
In research targeting the estimation of the parameters of a stochastic process model for a macroparasite population within a host, the team of Drovandi et al.~\cite{drovandi2011estimation} employed approximate Bayesian computation to model the immunity of the host as an unobserved model variable. 
Despite the very limited data, the authors had available, the process rate's time constants were inferred reasonably precisely with a grounded plausibility proof. The approach involved a three-stage Markov process for which the observed data likelihood was computationally intractable. The proposed algorithm was validated on an autologistic model prior to parameters inference from experimental data. Interestingly, the model also captured the extra-binomial variation of the immune system. 
The results were also supported by the study of Carr et al.~\cite{carr2021estimating} who presented a similar framework of Bayesian modelling and inference as in the work of Jorgensen et al.~\cite{jorgensen2022efficient}.
They proposed an efficient Bayesian inference method for a stochastic agent-based model. The study mitigated the use of the Bayesian setting 
\emph{(a)} by constructing lightweight surrogate models to substitute the simulations used for inference, and 
\emph{(b)} by circumventing the need for Bayesian sampling schemes and directly estimating the posterior distribution. 
This multi-staged approach demonstrated realistic results in tumour growth prediction.
Considering a similar scenario of tumour growth curve extraction, the work of Wang et al.~\cite{wang2022calibration} proposed a method for calibration of a Voronoi cell-based model for tumour growth using approximate Bayesian computation. 
Interestingly, the work involved as well estimating the distribution of parameters that govern cancer cell proliferation (i.e., the distribution of $\beta$ in Equation~\ref{eq1}) and recovering outputs that match the experimental data. 
Their results showed that the proposed approach, and its multi-stage extension, provided insights into tumour growth and a good quantification of this process uncertainty.
Multi-stage calibration describes a very promising avenue to explore, also because it is also supported by work that fuses mechanistic modelling and ML. A very good candidate subsuming these principles is the work of Axenie et al.~\cite{axenie2022fuzzy}, where the optimization step was reduced to a simple feed-forward inference through a logic model of spatio-temporal interactions of the agents. This modelling stage allowed the system to `inject' proper biases in the learning process of the mapping $F$ in the generic process captured by Equation~\ref{eq1} while keeping inference efficient. Additionally, the employed metrics of performance demonstrated a tight coupling among the micro- and macroscopic dynamics of the agents. Interestingly, the calibration of microscopic parameters took into account also the aforementioned scale coupling for a plausible candidate parameter configuration.

\subsection{Models comparison | Benchmarking}
\label{sec:Comparison_Benchmarking}

Given the availability of different computational techniques and approaches, the scientific community acknowledges the need to assess and compare different explanatory computational models. In particular, this abundance of \insilico{} models gives rise to the question of what criteria investigators should consider when making decisions of appropriateness. 
One commonly used process to address this question is model benchmarking. Model benchmarking stands for the assessment and quantification of \insilico{} models according to well-specified criteria. This provides the means to objectively infer metrics and uses these to compare the models, enabling a framework for improved selection.
Naturally, without proper standardization the benchmarking criteria can be strongly influenced by the research aims of the individual investigator(s), institutions or national funding policies, thus, causing biases and misconceptions. 
Here, we elaborate on the set of criteria that we believe to be crucial for ABM simulations in biomedicine and hence well-suited for benchmarking studies in the future.

Naturally, computational models' primary objective is to achieve a high-fidelity predictive performance. 
To accomplish this, specific measured data need to be reproduced by the \insilico{} model in a repeatable fashion (`replicability'). 
The capability in doing so is quantified using measures of `\emph{accuracy}.' 
In the biomedical context, such actions could relate to various anatomical, physiological, omics or other types of biological information. 
For instance, the study of Borlin et al.~\citep{Borlin_et_al:2014} presented an agent-based model that quantitatively captures several biological measurements of autophagy, which plays crucial roles in cellular and organismal homeostasis, including the response to diverse stresses. 
Ideally, the model performance refers to some type of agreement with such experimental data; for instance, this could be the number of cells or subcellular vesicles of a certain type. 
Nevertheless, oftentimes such agreement may be only of a qualitative nature, which can still exhibit significant explanatory power. For instance, a computational model that captures up- or down-regulation of certain metabolic pathways in the right context, has value, even if the magnitude of change is not quantitatively accounted for. 
However, in the presence of multiple plausible models, quantitative accuracy is a crucial factor and may determine which models to select over others.

As previously discussed (section \ref{sec:ML_assisted_Calibration}), computational models and in particular mechanistic models require parametrization using biological information. These should be ideally experimentally measured, and at least based upon plausible evidence.
Given that such parametrization may be difficult to conduct based on real-world data, it is best to minimize the presence of estimated model parameters while also reduce to the absolute necessary, based on the design of the agent-based model, the list of assumptions. 
Arguably models that require fewer parameters and less guesswork can generally be considered superior to those where significant subjective inputs are necessary. 
When multiple models confer equivalent agreement with experimental data, models with comparably reduced `\emph{complexity}' are preferable to others, in accordance with Occam's Razor \cite{domingos1999occams}.
Thus, with regard to mode benchmarking, it is important to consider and compare model complexities. 
At the very least, model complexity should reflect the complexity and disparity of the empirical data used to inform and calibrate the \insilico{}.
It is beyond the scope of this review however to give a comprehensive overview of model complexity measures, and the interested reader is recommended to read \citep{domingos1999occams, spiegelhalteretal2002bayesian}.

\emph{Explanatory power} is arguably the most important of all aspects of an \insilico{} model in biomedicine. 
A hypothetical model A, which is considered in this example as a black-box one, may perform significantly better in reproducing experimental data than a hypothetical mechanistic model B. 
However, the real-world value of model B may still be higher. 
No patient will agree to surgery without a proper explanation of the underlying reasoning. 
Indeed, model A's applicability is \emph{a-priori} limited in contexts where the stakes are high and interventions can have long-term implications. Nevertheless, to this date the explainable category of model B has seen very limited translation and thus application in the clinical setting. 

However, explanatory power is also difficult to quantify and so remains rarely assessed in existing benchmarking studies. 
A potential framework to assess and grade a model's explainability is as follows: 
in its most basic form, a model could be associated with a binary variable for being of the black-box or white-box type. 
In many cases, this could be extended to a spectrum of explainability, with the best models covering multiple data modalities comprising different spatial and temporal scales. 
Explainability is an active area of AI research and is particularly important to consider when assessing the usage of computational models in biomedical applications, such as for instance in cancer treatment \citep{Wijeratne_Vavourakis:2019, ventosoetal2020immuneresponse, cognoetal2022lungfibrosis, Hadjicharalambous_et_al:2022}.

Biomedical applications are in many regards very challenging, particularly when it comes to personalised medicine. Many diseases can progress very differently depending on lifestyle, age, genetics and comorbidities. In order for a computational model to have predictive power, it requires the capability to account for variation as observed in a given system (e.g., interpersonal variation, age-dependent changes, drug-induced changes, etc.). This aspect could therefore be accounted for in the benchmarking of computational models, for instance by producing distributions of outputs for multiple runs with different initial settings and/or stochastic dynamics. At the same time, computational models need to be also consistent, so that a given result does depend on very specific initial conditions. Such variation and consistency can be captured using `\emph{sensitivity}' analysis \citep{ten2016sensitivity, niidaetal2019sensitivity, Lima_et_al:2021}.

It is not astounding that most bench-marking studies focus on computational performance, which depends on `\emph{computational efficacy}.' 
Needless to say, the more efficiently and faster a model can be simulated, the better suited it is for parameter optimisation, refinement, and adaptation. 
While this does not necessarily entail enhanced explanatory power, it can lead to an edge when it comes to real-world applications. 
This is particularly true in the biomedical context where data acquisition has reached very high throughput levels, requiring frequent analysis, fast processing, and repeated updating and adaptation of computational models. Advanced code optimization techniques can significantly reduce computational time \citep{gonzalezetal2018optimization}. 
Moreover, ML methods often out-compete mechanistic, multi-scale models in terms of computational efficacy \citep{angioneetal2022surrogate}. 
Hence, incorporating ML techniques can be advantageous for certain applications, such as for the prognosis of patients who may suffer from time-critical issues \cite{axenie2020chimera}.

To conclude this section, a number of factors that are relevant for comparing computational models can be outlined (Table~\ref{tab:2}). 
These factors can serve as a stepping stone to formulate an ecosystem comprising the multitude of computational models, and establish their strengths and weaknesses in a systematic benchmarking approach. 
Naturally, the factors need to be considered on a case-by-case basis, and their relative importance varies according to context and the stakes at hand.

\begin{xltabular}{\textwidth}{|l|X|X|}
\caption{Benchmarking multi-scale computational models. Here, important criteria for benchmarking are shown, along with some example quantities that can be objectively measured and relevant references. Please note that this table is not exhaustive.}
\label{tab:2}
\\ \hline
\multicolumn{1}{|c|}{\textbf{Criterion}} & \multicolumn{1}{c|}{\textbf{Quantities}} & \multicolumn{1}{c|}{\textbf{References}}
\\ \hline 
\endfirsthead
\hline
\endlastfoot
Accuracy & Mean squared error (MSE), Root MSE (RMSE), F-score, Receiver operating characteristic (ROC) curve & \citep{Borlin_et_al:2014, Lima_et_al:2021} \\ \hline
Complexity & $\#$ of model parameters, model complicatedness, connectivity density of ontology graph, Bayesian complexity & \citep{mandes2017complexity, sunetal2016complicated}
\\ \hline
Explanatory power & accordance between emergent behaviour and experimental observations & \citep{glen2019agent, baueretal2021creative,Ruscone_et_al:2023}
\\ \hline
Model robustness/sensitivity & coefficient of variation, sensitivity indices & \citep{ten2016sensitivity, niidaetal2019sensitivity, Lima_et_al:2021,niidaetal2019sensitivity}
\\ \hline
Computational efficacy & Lines of code (LOC), run-time, memory usage, energy demand & \citep{gonzalezetal2018optimization, angioneetal2022surrogate}
\\ \hline
\end{xltabular}

\section{Discussion and outlook}

\subsection{Modern mechanistic modelling approaches in cancer biomedicine}

Agent-based models have proven to be invaluable tools for providing new insights and testing hypotheses on complex heterogeneous systems, including biological systems. 
In the context of cancer biomedicine, this novel mechanistic approach offers a unique way 
\emph{(a)} to fuse information from \invitro{} and \invivo{} data including for instance gene expression and medical imaging, 
\emph{(b)} to develop novel \insilico{} procedures for patient diagnosis and neoplasia stratification, 
\emph{(c)} to identify and compare the efficacy of potential treatment strategies as well as narrow the cardinality of the set of possible experiments down to the most promising ones while also 
\emph{(d)} to aid reducing animal testing and clinical trials on humans.

Recent efforts aimed at employing ABM to model tumour tissues have been presented in subsection~\ref{sec:Mechanistic_ABM}, which unveils the versatility of the approach. 
In fact, at multiple spatial and temporal scales, the models examined have demonstrated how various aspects of cancer dynamics can be captured, spanning from the pathogenesis to the growth and interaction with somatic cells. 
Moreover, while some 3D models can accurately replicate the morphological features of \invitro{} tumours and convey spatial characterisations, in others the agent-based compartment is seamlessly interfaced with different techniques.
This synergistic combination enables leveraging advantages of the different approaches, and addresses the general problem of many statistical methods that do not account for the mechanisms. Hence, agent-based models are excellently suited to produce explanatory powerful computational models that can integrate various data modalities, heterogeneous information as well as background information that may not be readily available in the form of experimental measurements.  

\subsection{Calibration `recipes' for validating agent-based simulations}

Calibration is the methodological procedure to compute model parameter configurations when the \insilico{} model can produce realistic simulation outputs. 
This process can be approached from multiple perspectives and there is a plethora of tools capable to drive `optimizing' the parameters and, thus, to improving the fidelity of large-scale agent-based models in cancer biomedicine. 
Independent of their inner workings, the calibration process is a `search' in a large parameter space. 
This search can be guided (as in the case of machine learning approaches) by embedding information about the physics of the process being modelled or rules of the evolution of the process dynamics. 
While most of the calibration approaches formulate the `search' as an optimization method (i.e., to minimize a distance among some performance indices), there is much more to gain from embedding rules and physics in the optimization process. 
This has been formally described in subsection~\ref{sec:Calibration}, where a comparative benchmark has been developed to offer the readers a `recipe' for approaching the validation of agent-based models through calibration. 
However, we acknowledge that there is no `one size fits all' solution, and we thus recommend formulating the calibration process as suggested in subsection~\ref{sec:Fusing_Models} followed by a benchmarking as suggested in subsection~\ref{sec:MachineLearning_ABM} and in Table~\ref{tab:2}. 
Besides, choosing the best fitting candidate from the ``models zoo'' is thoroughly covered in subsection~\ref{sec:MachineLearning_ABM} and in the excellent review of Metzcar and his colleagues \cite{Metzcar_et_al:2019}. 
Starting there the reader could get a grasp on the necessary functionality and capabilities of already successful simulation tools and machine learning and physics-informed learning algorithms in cancer biomedicine.

\subsection{Outlook to `non-invasive' calibration of agent-based models}

While a computer simulation is a cost-effective and safe way to evaluate hypotheses on new cancer therapies, immune--tumour interactions, and bio-physics of tumour invasion, existing simulators are often impractical due to inefficient control interfaces. 
This hinders the data exchange of the simulation evolution with the `external' world where signals, data, or decisions to update the temporal trajectory of the simulation might be applied.
A mature simulator, under the control of external systems, can be used to run the models being tested within a closed-loop environment. 
However, state-of-the-art simulators are often impractical for the task at hand due to their inefficient control interfaces. 
The challenges can be divided into two primary areas: the requirement for efficient synchronization, and the need for flexible and scalable data processing during runtime.
In terms of the synchronization model, state-of-the-art ABM simulators primarily use a conservative step-based approach, which needs synchronization at every interaction point. 
However, the high frequency of interactions caused by data collection can result in an overwhelming amount of message flow, leading to significant overhead.
Alternatively, an optimistic approach that allows for full asynchronicity and causality violations exists, but implementing a rollback mechanism to address missing decision-making or data collection times is complex, error-prone, and not supported by most ABM simulators.
Processing simulation data during runtime poses other challenges, e.g., storage and large volumes of simulation outputs, secure and fast transfer of sensitive clinical data, extracting valuable insights from a limited data amount, making predictions based on an unevenly distributed sample etc.
The objective is to dynamically instruct the running simulation, allowing it to store and output the data of interest while avoiding any redundancy. 
An essential yet challenging aspect of this problem is performing temporal operations and on-the-fly data processing within queries, for example, filtering or computing data to retrieve results over a specific time interval.

\section*{Acknowledgements}

V.V. acknowledges the financial support from the Cyprus Cancer Research Institute, as part of the project ``PROTEAS'' (CCRI Bridges in Research Excellence; Funding Agreement No.~CCRI\_2021\_FA\_LE\_105).

\bibliographystyle{tfnlm}
\bibliography{bibliography}

\end{document}